\documentclass[aps,prb,reprint,amsmath,amssymb,superscriptaddress,longbibliography]{revtex4-2}

\newcommand{\NBCO}{Na$_2$BaCo(PO$_4$)$_2$\xspace}

\usepackage{comment}
\usepackage{graphicx}
\usepackage{dcolumn}
\usepackage{bm}
\usepackage{amsmath}
\usepackage{amssymb}
\usepackage{xcolor}
\usepackage{xspace}
\usepackage{xfrac}
\usepackage{color,soul}
\usepackage{graphicx}
\usepackage[colorlinks=true,urlcolor=blue,citecolor=blue,linkcolor=blue,bookmarks=false,pdfstartview={FitH}]{hyperref}

\begin{document}
	
	\title{Field-tuned incommensurate fan phase in an Ising-like triangular antiferromagnet}
	
	\author{D.~Flavi\'{a}n}
	\affiliation{Clarendon Laboratory, University of Oxford Physics Department, Parks Road, Oxford OX1 3PU, UK}
	
	\author{R.~Okuma}
	\affiliation{Clarendon Laboratory, University of Oxford Physics Department, Parks Road, Oxford OX1 3PU, UK}
	\affiliation{Institute for Solid State Physics, University of Tokyo, Kashiwa, Chiba 277-8581, Japan}
	
	\author{P.~Manuel}
	\affiliation{ISIS Facility, Rutherford Appleton Laboratory, Chilton, Didcot OX11 0QX, UK}
	
	\author{Q.~Huang}
	\affiliation{Department of Physics and Astronomy, University of Tennessee, Knoxville, Tennessee 37996, USA}
	
	\author{H.~Zhou}
	\affiliation{Department of Physics and Astronomy, University of Tennessee, Knoxville, Tennessee 37996, USA}
	
	\author{R.~Coldea}
	\affiliation{Clarendon Laboratory, University of Oxford Physics Department, Parks Road, Oxford OX1 3PU, UK}

	\begin{abstract}
	Using single-crystal neutron diffraction we report the observation of a field-tuned incommensurate phase in a field range just below magnetization saturation in the Ising-like triangular-lattice antiferromagnet \NBCO. This phase occurs for magnetic fields applied along the Ising axis and exhibits an in-plane incommensurate propagation vector that moves along the hexagonal Brillouin zone boundary upon varying field, challenging the simple XXZ antiferromagnetic description of the system. Within a semiclassical framework, we successfully identify this pre-saturation phase as a coplanar fan phase and study the role of dipole-dipole interactions and bond-dependent exchange anisotropy in its stabilization mechanism. We additionally present a minimal exchange model including interlayer couplings, which reproduces the complex magnetic phase diagram in field including the observed field-dependence of the incommensurate propagation vector in the fan phase.
	\end{abstract}

    \date{\today}
	\maketitle
    
\begin{figure}
	\centering
	\includegraphics[scale=1]{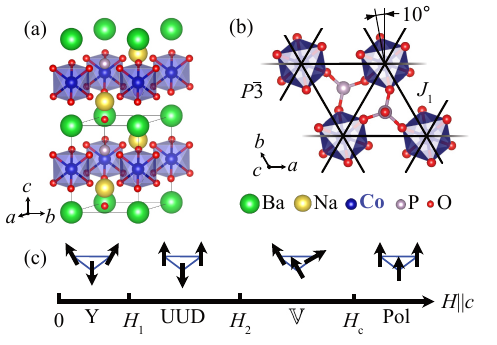}
	\caption{(a) \NBCO crystallizes in a $P\bar{3}$ structure, where magnetic Co$^{2+}$ ions sit at the center of oxygen octahedra forming a triangular lattice. (b) A rotation of the octahedra around the \textit{c} axis breaks mirror planes normal to the $a$ and $b$-axes, impacting the allowed exchange pathways for neighbors in adjacent triangular layers. Black bonds (labeled $J_1$) indicate the nearest-neighbor couplings, relevant for Eq.~(\ref{Eq:IntroH}) and Eq.~(\ref{eq:TLAFM_Full}). (c) Schematic of the expected magnetic phase diagram for the triangular-lattice XXZ model in Eq.~(\ref{Eq:IntroH}) in axial magnetic field ($H \| z$). The additional phase found in Fig.~\ref{fig:NBCO_Fan_diffraction} in-between the $\mathbb{V}$ and the field-polarized phase calls for a refinement of the spin Hamiltonian beyond Eq.~(\ref{Eq:IntroH}).}
	\label{fig:Introduction}
\end{figure}

Frustrated quantum magnets provide a key platform to explore experimentally novel phases of matter stabilized by strongly competing interactions. In such systems, magnetic anisotropies often play a critical role in the ground state selection mechanism, by breaking symmetries and/or lifting degeneracies. The nearest-neighbor triangular-lattice spin-1/2 easy-axis XXZ antiferromagnet with spin Hamiltonian 
\begin{equation}
\mathcal{H}_{0} = J_{zz} \sum_{\left<ij\right>}  [S^z_i S^z_j + \Delta(S^x_i S^x_j + S^y_i S^y_j)] -g_{zz}\mu_{\rm B}\mu_0 H\sum_i S_i^z
\label{Eq:IntroH}
\end{equation}
with $J_{zz}>0$ and $0<\Delta<1$ has served as a cornerstone of anisotropic frustrated magnetism. This model has been predicted to have a characteristic sequence of distinct three-sublattice magnetically ordered phases as a function of field $H$ applied along the Ising axis \cite{nishimori1986magnetization,chubukov1991quantum,YamamotoXXZPhasesAxial} (see Fig.~\ref{fig:Introduction}(c)), the so-called Y-phase, up-up-down (UUD) 1/3$^{\rm rd}$ magnetization plateau, and $\mathbb{V}$-phase, before transition to full spin polarization.  

\NBCO stands out as a rare experimental realization of this model in the magnetically clean limit of no structural disorder \cite{zhong2019strong,li2020possible,sheng2022two,gao2022spin}, with octahedrally coordinated Co$^{2+}$ ions arranged in a triangular lattice (Fig.~\ref{Eq:IntroH}). The experimental observation via inelastic neutron scattering (INS) measurements of a broad continuum of excitations in zero field co-existing with long-range magnetic order Bragg peaks  \cite{GaoNBCORoton,sheng2025continuum,NBCO_Leonie} has sparked much theoretical interest in relation to a possible proximity to a quantum spin liquid phase \cite{GaoNBCORoton,sheng2025continuum}, or a novel mechanism of lifting the continuous classical degeneracy of the XXZ model by frustrated interlayer couplings \cite{NBCO_Leonie}. The INS measurements in the field-polarized phase showed sharp magnon dispersions as expected for this phase and their analysis established that the dominant interactions are indeed the nearest-neighbor XXZ exchanges in Eq.~(\ref{Eq:IntroH}), with negligible further neighbor couplings \cite{sheng2022two,NBCO_Leonie}. 

However, subtler forms of anisotropy, such as bond-dependent exchange terms and dipolar couplings, can have significant effects in frustrated systems \cite{d7Kitaev,d7KitaevII,d7KitaevIII}. By breaking residual spin symmetries, these terms can shift the location of soft modes in momentum space and promote incommensurate or otherwise unconventional magnetic order \cite{shiba1982incommensurate,Li2016TLFAM,KitevZ2Vortex_2, KitevZ2Vortex_2}, effects that may remain largely invisible in spectroscopic probes but become evident in the structure of ordered phases.

Here we address this point by presenting a series of magnetic neutron diffraction experiments on a single-crystal sample of \NBCO, monitoring the evolution of magnetic phases under fields applied along the easy axis ($H\|c$). Besides magnetically ordered phases with Bragg peaks with in-plane propagation vector at the 2D Brillouin zone corner K-point associated with in-plane 3-sublattice order (as expected for a pure XXZ model, see Fig.~\ref{fig:Introduction}(c)), we find clear evidence for an additional novel phase in an extended field range pre-saturation, characterized by a field-dependent incommensurate in-plane propagation vector that moves in field away from the K-point along the Brillouin zone boundary towards the M point. We discuss several scenarios for the origin of this phase and argue that often-neglected dipolar couplings provide a natural mechanism for its stabilization. Finally, we address the weak but relevant inter-layer couplings and their role in stabilizing a complex evolution of the out-of-plane magnetic propagation vector. Based on knowledge of the propagation vectors as a function of magnetic field strength, we propose a refined spin Hamiltonian for the strongly frustrated \NBCO that captures all key features of its phase diagram for axial fields. 

The remainder of this paper is organized as follows. Section~\ref{S:methods} covers the applied methodology, followed by Section~\ref{S:diffraction}, which describes the neutron diffraction results in detail. In Section~\ref{S:theory}, a theoretical framework is built to understand the diffraction data, including bond-dependent exchange anisotropies, dipole-dipole interactions, and interlayer couplings. Finally, Section~\ref{S:discussion} discusses the results of our study within a wider scope, and Section~\ref{S:conclusions} summarizes the conclusions.  
		
	\section{Methods}
    \label{S:methods}
	The single-crystal sample of \NBCO used for this study was grown using a NaCl flux method as detailed elsewhere \cite{zhong2019strong}. \NBCO crystallizes in the $P\bar{3}$ space group, with lattice parameters $a$ = 5.312 \AA~and $c$ = 6.983 \AA~ at room temperature \cite{NBCO_Leonie}. Large samples (mm-size) tend to contain two twins related by a 180$^\circ$ rotation around the $c$-axis \cite{NBCO_Leonie}. The diffraction patterns of the two twins differ only in the relative intensities of some of the reflections. Analysis of the magnetic diffraction pattern from the $\sim 70$~mg crystal used in the magnetic diffraction experiments to be discussed later suggests that it likely has an unequal structural domain population. 
		
	Neutron diffraction data were collected using the time-of-flight diffractometer WISH \cite{chapon2011wish} at the ISIS Facility in the UK. A $^3$He-$^4$He dilution refrigerator with a base temperature of 0.05~K was used in combination with a 10~T vertical cryomagnet to access the full phase diagram up to magnetic saturation above 1.65~T. The sample was attached using GE varnish on a copper sample holder to provide good thermalization, with the $c$-axis vertical. Wavevectors throughout the text are given in reciprocal lattice units of the hexagonal unit cell. Diffraction maps were collected at a number of magnetic fields and temperatures, and the data was reduced and converted to diffraction intensity as a function of wavevector using the Mantid \cite{Mantid} and MSlice \cite{MSlice} software suites. Measurements at 2.5~T deep in the field-polarized phase are used as an estimate of the non-magnetic diffraction background signal near the positions of the magnetic order Bragg peaks, and this data was subtracted from all lower-field diffraction patterns collected in the same crystal orientation.

    Semiclassical Linear Spin Wave Theory (LSWT) calculations have been performed using the \textsc{sunny}.jl (v0.7.2) package \cite{Sunny2025}. Long-ranged dipolar interactions have been included in the calculations using the Ewald summation method as implemented in the package. 

\begin{figure}[h!]
	\centering
	\includegraphics[scale=1]{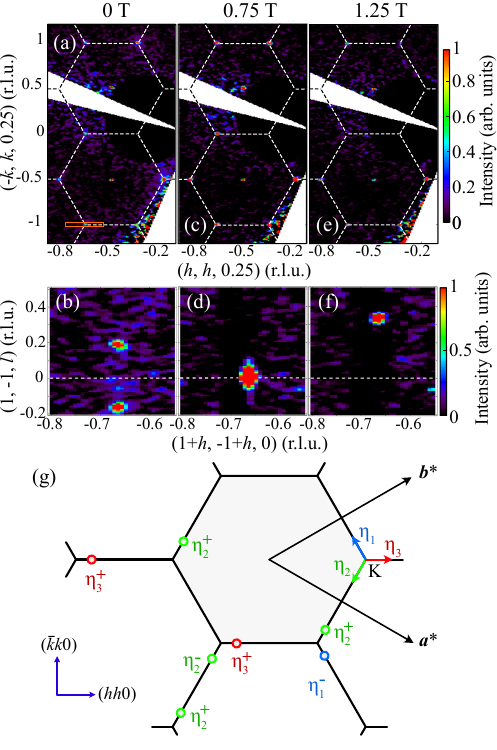}
	\caption{(a-f) Representative single crystal neutron diffraction maps in phases I-III at three selected magnetic fields, increasing from left to right. Dashed lines in top row panels show boundaries of structural Brillouin zones. Note all magnetic Bragg peaks occur at the two-dimensional Brillouin zone corner K points in (a), (c), and (e), but with different out-of-plane component $\pm q_z$ in the three phases, (b) $q_z$ near 1/6 in zero field (phase I), (d) $q_z=0$ at 0.75~T in the 1/3$^{\rm rd}$ magnetization plateau UUD phase II, (f) $q_z$ near 1/3 at 1.25~T in phase III. Note that in (a), (c), and (e) data is integrated in a wide range in the interlayer direction $0\leq l\leq0.5$ in order to capture both structural and magnetic peaks in the same frame. The transverse wavevector integration range in (b), (d),  and (e) is shown with an orange rectangle in the upper left corner of (a), and the dashed horizontal line highlights $l=0$. The color scale is shared across same row panels. (g) Sketch of the two-dimensional reciprocal space (first Brillouin zone shaded), showing the geometry of the experiment. Arrows around a K point (bottom right, blue/green/red) show the definition of the in-plane $\bm{\eta}_{i}$ ($i=1-3$) magnetic propagation vectors in phase IV and colored-coded circles indicate corresponding magnetic Bragg reflections observed experimentally in this phase (see Fig.~\ref{fig:NBCO_Fan_diffraction}), with an exaggerated deviation from their respective K point, for clarity. $\pm$ superscripts indicate the sign of the out-of-plane magnetic Bragg peak components.}
	\label{fig:Diffraction_YUV}
\end{figure}

\section{Neutron diffraction results}
\label{S:diffraction}
\begin{figure*} 
	\centering
	\includegraphics[scale=1]{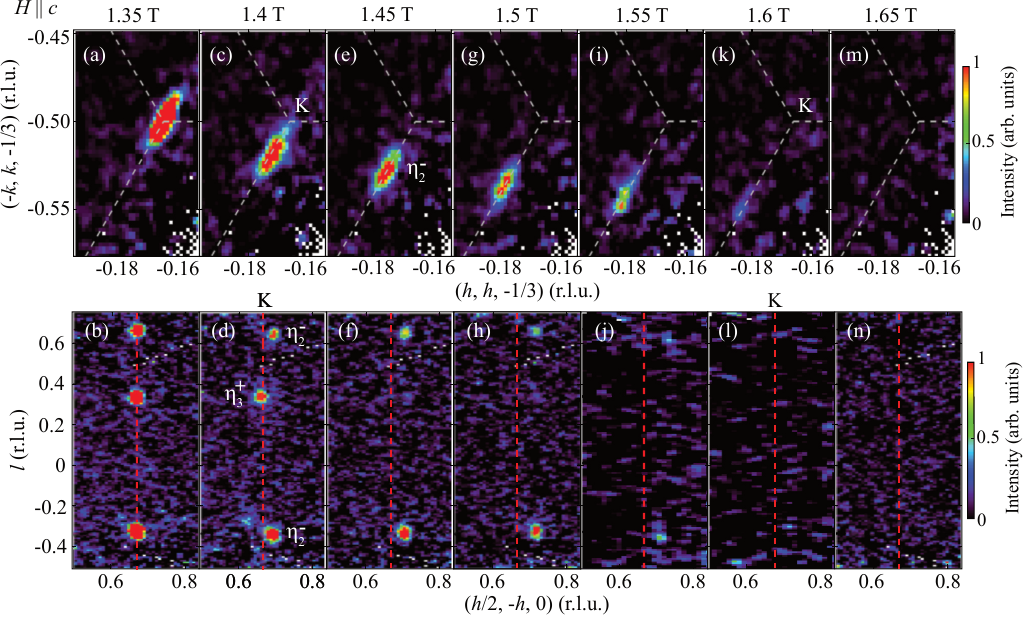}
	\caption{Detail of the magnetic neutron diffraction in phase IV as a function of magnetic field increasing from left to right. (Upper panels) Magnetic diffraction in the $(h,h,-1/3)\times(-k,k,-1/3)$ plane integrated for -0.35 $< l <$ -0.30 following the evolution of a magnetic reflection with $\bm{\eta}_2^-$ propagation vector. Dashed lines show the 2D Brillouin zone boundaries. Upon increasing field, the magnetic Bragg peak moves away from the K point along one of the three Brillouin zone boundaries towards an M point. (Lower panels) Magnetic diffraction in the orthogonal plane $(1,-2,0)\times(0,0,1)$, where the dashed vertical red lines indicate the in-plane K point position for reference. The color scale is shared among the figures. Panel (d) shows Bragg peaks corresponding to two different propagation vectors $\bm{\eta}_2^-$ ($l$ near $-1/3$ and 2/3) and $\bm{\eta}_3^+$ ($l$ near $1/3$). The intensity of the Bragg reflections is reduced as the spins get progressively more polarized by the field, and disappears completely above the critical field $\mu_0H_c=1.65$~T.}
	\label{fig:NBCO_Fan_diffraction}
\end{figure*} 

Neutron diffraction measurements at a base temperature of 0.065~K reveal four  distinct magnetic phases (I-IV) as a function of increasing $c$-axis field. Representative diffraction patters in phases I-III are shown in  Fig.~\ref{fig:Diffraction_YUV}. Consistent with previous reports \cite{sheng2022two,xiang2024giant}, prominent magnetic reflections are observed with propagation vectors of the form ${\bf Q}^{\pm}_{\rm I-III}=(1/3,1/3,\pm q_z)$, with a different $q_z$ in the three phases. The in-plane (1/3,1/3) propagation vector corresponds to three-sublattice order in each layer and a finite out-of-plane propagation vector $q_z$ indicates a phase offset between stacked layers. The observed magnetic reflections are sharp in all three directions, as expected for long-range magnetic order, and disappear upon heating, confirming their magnetic character. In Fig.~\ref{fig:Diffraction_YUV}(a,b) (zero field), the interlayer propagation vector is incommensurate at $q_z=0.170(1)$. In Fig.~\ref{fig:Diffraction_YUV}(c,d) at 0.75~T deep in the 1/3$^{\rm rd}$ magnetization plateau phase $q_z=0$, i.e. all layers have the same up-up-down structure. In Fig.~\ref{fig:Diffraction_YUV}(e,f), at 1.25~T in phase III, $q_z=0.328(2)$. 
Data at higher fields above 1.35~T reveal magnetic Bragg peaks associated with a distinct magnetic phase IV. Fig.~\ref{fig:NBCO_Fan_diffraction} (top row) focuses on the field-dependence of one such magnetic Bragg peak: as the field increases, it moves in reciprocal space away from the two-dimensional Brillouin zone corner K-point along the Brillouin zone boundary towards the M point. In particular, this reflection is found at $\mathbf{Q}^{-}_{\rm IV} = (1/3+\delta , -2/3-2\delta, -0.328(2))$, where $\delta$ evolves continuously from 0 at 1.35~T to a maximum $\delta_c$ = 0.027(2) at the critical field $\mu_0H_c=1.65$~T, above which all spontaneous order magnetic Bragg peaks disappear, which we attribute to the transition to the field-polarized phase. The field dependence of the incommensuration, $\delta$, is shown in Fig.~\ref{fig:Fan_model}(d), where it is compared to a theoretical model as detailed later in Sec.~\ref{S:theory}. 
 
We are able to observe several magnetic Bragg peaks in phase IV. To simplify the notation, we define the following three propagation vectors based on their deviation from the Brillouin zone corner K-point, i.e., ${\mathbf Q}_{i,{\rm IV}}^{\pm}$ = (1/3,1/3,0) + $\bm{\eta}_i^\pm$, where $\bm{\eta}_1^\pm = (-2\delta ,\delta ,\pm q_z)$, $\bm{\eta}_2^\pm = (\delta ,-2\delta ,\pm q_z)$ and $\bm{\eta}_3^\pm = (\delta ,\delta ,\pm q_z)$ are related by $120^\circ$ rotation around the $z$-axis, with $q_z = 0.328(2)$. The in-plane projections of these three vectors are shown in Fig.~\ref{fig:Diffraction_YUV}(g) (blue/green/red arrows, respectively) for reference. Following this notation, the upper panels in Fig.~\ref{fig:NBCO_Fan_diffraction} show the field evolution of a reflection with propagation vector $\bm{\eta}_2^-$.  We summarize all the observed incommensurate peaks in Fig.~\ref{fig:Diffraction_YUV}(g), where several inequivalent propagation vectors were detected. Importantly, a magnetic Bragg peak in phase III associated with propagation vector $\mathbf{Q}_{\rm III}^{\xi}$ ($\xi=\pm$) does not split into three magnetic peaks corresponding to $\bm{\eta}^\xi_i$ ($i=1-3$) in phase IV, but evolves continuously into one and only one of those, suggesting that the latter is a single-$\mathbf{Q}$ rather than a triple-$\mathbf{Q}$ phase. We associate the co-existence of magnetic Bragg peaks with $\pm q_z$ components to the presence of magnetic domains associated with different structural domains related by $180^{\circ}$ rotation around their $c$-axis, commonly found in \NBCO crystals of a size comparable or larger than the sample studied in our experiments \cite{NBCO_Leonie}. Importantly, the magnetic peaks of type $\bm{\eta}_2^-$ are the strongest overall, suggesting that they are likely associated with the majority structural domain. 

As a final remark, at each investigated Brillouin zone corner, we observe incommensurate peaks with propagation vector along different $\bm{\eta}_i$ directions for $+q_z$ and $-q_z$ as indicated in Fig.~\ref{fig:Diffraction_YUV}(g). One such instance is clearly displayed in Fig.~\ref{fig:NBCO_Fan_diffraction}(d), where reflections corresponding to $\bm{\eta}_2^-$ and $\bm{\eta}_3^+$ are observed. Note that as the $\bm{\eta}_3^+$ reflection (near $l=+1/3$) shifts out of the integration range upon increasing field its intensity appears to vanish at a much lower field than the two $\bm{\eta}_2^-$ reflections (near $l=-1/3,2/3$), which can still be clearly visible in panels (f,h); however, by monitoring the intensity in the $hk$ plane near $l=\pm1/3$ and $2/3$ (not shown) we have confirmed that all magnetic peaks (including the above three) disappear at the same critical field within experimental uncertainty.   

\section{Beyond the XXZ model}
\label{S:theory}	

Extensive inelastic neutron scattering experiments have been devoted to determine the dominant spin interactions in \NBCO. These studies have analyzed spin waves in the spin polarized state under strong magnetic fields, both parallel \cite{sheng2022two,GaoNBCORoton,sheng2025continuum} and transverse \cite{NBCO_Leonie} to the easy-axis direction (crystallographic $c$ axis).  A model of nearest-neighbor interactions between effective spin $S = 1/2$ Co$^{2+}$ magnetic moments with an easy-axis XXZ-type exchange anisotropy and easy-axis $g$-tensor, reproduces the magnetic excitation spectrum exceptionally well for both $H\| c$ and $H\| b^*$ \cite{NBCO_Leonie}. Nonetheless, several qualitative features of our detailed diffraction data cannot be explained within this model.  First, the existence of a $q_z\neq0$ component for some ordered phases must be attributed to small, but nonetheless non-negligible, interlayer interactions.  The magnitude of such couplings is bounded by the out of plane dispersion probed by INS and shown to be at most a few percent of the main $J_{zz}$ exchange \cite{NBCO_Leonie}. Second and crucially, the existence of the additional incommensurate pre-saturation phase shows that a more complex Hamiltonian is needed to describe the physics of \NBCO, beyond the dominant XXZ nearest-neighbor model in Eq.~(\ref{Eq:IntroH}). 

In the following, we present an extended exchange model that captures these novel features. The analysis is based on extending the minimal model in Ref.~\cite{NBCO_Leonie}, given in Table~\ref{table:Full_H}.  Note that we do not aim at refining the main exchanges of the proposed spin Hamiltonian in Eq.~(\ref{Eq:IntroH}),  but to motivate the additional interaction terms needed to capture the magnetic diffraction results.

\subsection{Magnons in the polarized phase}

The existence of incommensurate magnetic structures is generally regarded as a signature of competing interactions. As a first step, we explore a series of simple scenarios that would result in incommensurate structures in a triangular lattice antiferromagnet.  The simplest case is to consider further-neighbor couplings, in the form of a $J_1-J_2-J_3$ model, while keeping the XXZ character for $J_1$. In the easy-plane model,  incommensurate phases have been predicted for certain $J_2/J_1$ and $J_3/J_1$ ratios \cite{ivanov1995spiral,GongSpiralsj1j2j3}. However, rather large further-neighbor couplings would be required to stabilize states with the propagation vector as observed in phase IV, and such values are ruled out by the inelastic spectrum in a polarising field \cite{sheng2022two,GaoNBCORoton,sheng2025continuum,NBCO_Leonie}.

Planar incommensurate structures appear close to saturation in the distorted, spatially anisotropic $J$-$J^\prime$ triangular lattice model, for any $J>J^\prime$ \cite{StarykhTLAFM_Saturation,IncommJJPrimeBalents}. Such anisotropy can be triggered by structural distortions of the triangular planes. However, no evidence for such a structural distortion breaking the three-fold symmetry is evident in the diffraction data either at zero or applied field, ruling out such a scenario in the present case. 

A more complex mechanism to stabilise incommensurate structures relies on the presence of Kitaev interactions \cite{KitevZ2Vortex,KitevZ2Vortex_2}. A non-zero Kitaev exchange (\textit{K}) explicitly breaks the U(1) symmetry and leads to a zero-field destabilization of the 120$^\circ$ structure in the otherwise pure Heisenberg Hamiltonian ($\Delta =1$).  The result is a $\mathbb{Z}_2$ vortex crystal, giving rise to a triple-$\bf{Q}$ structure with Bragg peaks appearing simultaneously along the three K to M directions, for $K<0$. The precise distance between vortex cores (and thus the value of the incommensurate parameter $\delta$) depends on the ratio of Kitaev to Heisenberg coupling, $-K/J_{zz}$. The presence of Kitaev-type exchange in \NBCO was considered in \cite{NBCO_Leonie}, concluding that if present at all, $K$ must be less than 5\% of the main $J_{zz}$ exchange, overall in agreement with the small value of $\delta$ in Fig.~\ref{fig:NBCO_Fan_diffraction}. However, the Kitaev-Heisenberg model would predict incommensurate order in the layers in zero field, which is not observed, and a triple-$\mathbf{Q}$ structure, whereas our experiments find evidence for multiple domains of single-$\mathbf{Q}$ structures.

\begin{figure}[tbp] 
		\centering
		\includegraphics[scale=1]{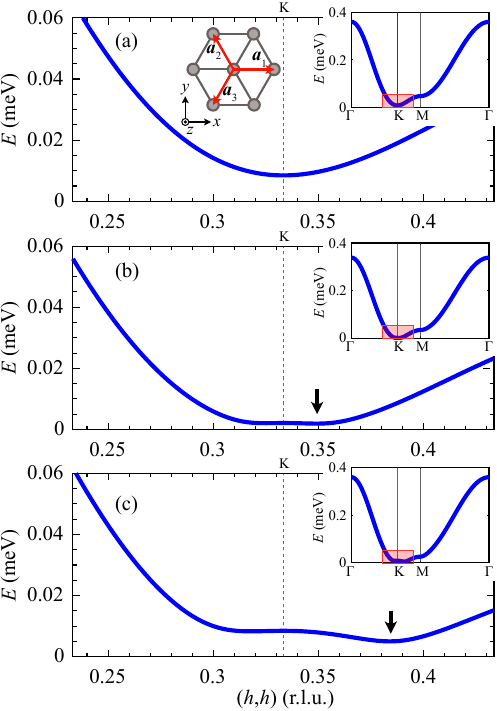}
		\caption{Comparing the location of the minima of the magnon dispersion in the field polarized state just above the critical field $H_c$ for different single-layer Hamiltonians. (a) In the pure XXZ model the branch minimum is at the K point (vertical dashed line). Inset shows the $\mathbf{a}_i$ vectors (red arrows) used in the definition of the anisotropic exchange terms in Eq.~(\ref{eq:TLAFM_Dispersion}). (b) Including the nearest-neighbor part of the dipolar interactions splits the K-point minimum into two inequivalent minima, the lowest of which is towards the M point (as indicated by a vertical black arrow). (c) Including the long-ranged dipolar couplings for the whole triangular layer further accentuates this splitting, but the qualitative results remain. Insets in the upper right corner in all cases show the whole magnon branch along high symmetry directions, while a red box shows the zoomed-in area of the main panels.}
		\label{fig:SWT_NN_Dip}
\end{figure}

Having considered and ruled out the above scenarios, let us explore how an often-neglected interaction can explain all our experimental observations: the dipolar couplings.  Importantly, this interaction can be regarded as a small source of bond-dependent exchange anisotropy and we will motivate our results from this perspective. Considering the small nearest-neighbor exchange energy in \NBCO together with the large $g$-tensor values (see Table~\ref{table:Full_H}) and a relatively small Co-Co distance, it is natural that this dipole-dipole coupling may play a role in the selection of the ground state. We define the dipole-dipole coupling as:
\begin{equation}
	\mathcal{H}_{\rm d} = -\sum_{\langle ij \rangle}\frac{\mu_0}{4\pi|{{\mathbf{r}_{ij}}}|^3} \left[ 3 (\mathbf{m}_i\cdot\mathbf{\hat{r}}_{ij})  (\mathbf{m}_j\cdot \mathbf{\hat{r}}_{ij}) - \mathbf{m}_i\cdot\mathbf{m}_j  \right],
\label{eq:DipolarCoupling}
\end{equation}
with the sum extending over all spin pairs $\langle ij \rangle$ in the lattice counted once, $\mathbf{m}_i = \mu_{\rm B} \tilde{g} \mathbf{S}_i$ is the magnetic moment at site $i$, $\mathbf{r}_{ij}$ is the vector linking ions at sites $i$ and $j$, $\mathbf{\hat{r}}_{ij}$ a unit vector along this direction, and $\tilde{g}$ the anisotropic $g$-tensor, which in the $xyz$ frame of Fig.~\ref{fig:SWT_NN_Dip}(a) is diagonal with entries $g_{xx}=g_{yy} \neq g_{zz}$ given in Table~\ref{table:Full_H}.

The magnetic order just below the transition to spin polarization at the critical field $H_c$ is directly related to the wavefunction of the magnon that closes gap upon lowering field at $H_c$, thus we may gain insight into the magnetic structure just below $H_c$ by studying the magnon dispersion in the polarized state. We model the magnon dispersion using semiclassical Linear Spin Wave Theory (LSWT), which is expected to be asymptotically exact in the limit of high fields in the field-polarized phase.  In the pure XXZ case, magnons condense at the Brillouin zone corner K-points, giving rise to the well-known $\mathbb{V}$ structure illustrated in Fig.~\ref{fig:Introduction}(c). The interaction in Eq.~(\ref{eq:DipolarCoupling}) is long ranged, and thus complex to treat analytically within LSWT. Yet, as a first approximation, we may consider only its nearest-neighbor part, which already captures qualitatively the key phenomenology. The dipolar coupling between nearest-neighbor (nn) $ij$ sites linked by the $\mathbf{a}_1$ bond can be written in the $xyz$ Cartesian frame (see inset in Fig.~\ref{fig:SWT_NN_Dip}(a)) as
\begin{equation}
\mathcal{H}_{{\rm d},\mathbf{a}_1} = K_{\rm d,nn}
\begin{pmatrix}
S_i^x \,\, S_i^y \,\, S_i^z\\
\end{pmatrix}
\begin{pmatrix}
-2g_{xx}^2 & 0 & 0\\
0 & g_{xx}^2 & 0 \\
0 & 0 & g_{zz}^2 \\
\end{pmatrix}
\begin{pmatrix}
S_j^x\\
S_j^y\\
S_j^z\\
\end{pmatrix}
\label{eq:NN_dipolar}
\end{equation}
with $K_{\rm d,nn} = \frac{\mu_0\mu_B^2}{4\pi a^3}\approx 0.36~\mu$eV. This clearly shows the bond-dependent character of the interaction that breaks the U(1) rotational symmetry of the system explicitly.  Note that due to the large $g$-tensor the energy scale of dipolar interactions $K_{\rm d,nn}g_{zz}^2\sim8\mu$eV becomes relevant relative to the main exchange $J_{zz}$= 122.5~$\mu$eV. Fig.~\ref{fig:SWT_NN_Dip}(b) shows the calculated magnon dispersion in the polarized phase along high symmetry directions in the vicinity of the K point. Crucially, already at the nearest neighbor level, the dipolar couplings split the minimum at the K point into two inequivalent minima along the ($hh$) direction. This is a direct consequence of breaking the U(1) symmetry (as for other couplings that have the same effect \cite{KitevZ2Vortex,KitevZ2Vortex_2}), and applies regardless of the magnitude of $K_{\rm d,nn}$ (as long as it is small compared to $J_{zz}$). The deeper minimum, and thus the one that closes gap at $H_c$, is displaced from the K point towards the M point, consistent with our experimental observations  of the location of the magnetic Bragg peaks just below $H_c$. For completeness, the calculation is extended to include the long-range dipolar couplings for the whole triangular layer, and the results are shown in Fig.~\ref{fig:SWT_NN_Dip}(c). A more accentuated splitting of the dispersion minimum is found in this case, but the qualitative result remains the same: dipolar couplings lead to gap closing at an incommensurate wave vector away from K along the K-M line. 

The coupling in Eq.~(\ref{eq:NN_dipolar}) is none other than a special form of bond-dependent anisotropic exchange on the triangular lattice. The general form of the nearest-neighbor exchange deduced using symmetry arguments is \cite{MaximovAnisotropicTLAFM}
\begin{equation}
\mathcal{H}_{1} = \mathcal{H}_{0} + \mathcal{H}_{\rm bd}, 
\label{eq:TLAFM_Full}
\end{equation}
\begin{multline}
	  \mathcal{H}_{\rm bd} = \sum_{\left<ij\right>} J_{\pm\pm}(\gamma_{ij} S^+_i S^+_j + \gamma_{ij}^* S^-_i S^-_j) \\
	- \frac{ iJ_{z\pm}}{2}[(\gamma_{ij}^* S^+_i - \gamma_{ij} S^-_i ) S^z_j + S^z_i (\gamma_{ij}^* S^+_j - \gamma_{ij} S^-_j )]\nonumber
\end{multline}
with the bond-dependent term $\gamma_{ij} = 1,  e^{i \frac{2\pi}{3}}, e^{-i \frac{2\pi}{3}}$, for bonds along the $\mathbf{a}_1$, $\mathbf{a}_2$, and $\mathbf{a}_3$ directions as defined in Fig.~\ref{fig:SWT_NN_Dip}(a), respectively. The above form is appropriate for an idealized crystal structure of \NBCO with space group $P\bar{3}m1$ with vertical mirror planes that bisect each Co-Co bond. The actual crystal structure is lower symmetry with no mirror planes (space group $P\bar{3}$), which allows for two additional nearest-neighbor exchange terms that we neglect in the following analysis; for more details, see Appendix~\ref{A}.

\begin{figure} 
		\centering
		\includegraphics[scale=1]{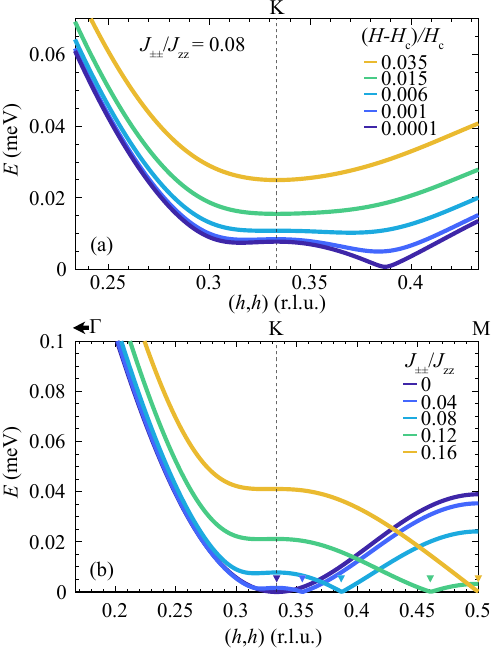}
		\caption{Magnon dispersion in the polarized phase in the presence of bond-dependent exchange anisotropy $J_{\pm\pm}$ as per Eq.~(\ref{eq:TLAFM_Dispersion}). Note the dispersion depends only on the magnitude and not the sign of the $J_{\pm\pm}$ term. (a) Dispersion as a function of field (color coded) at fixed $J_{\pm\pm}$. The minimum shifts away from the K point (dashed line) towards an incommensurate wavevector on the K-M line when the applied field is sufficiently close to $H_c$, i.e., when the spin gap becomes small enough to be comparable to the strength of the anisotropic exchange. (b) Dispersion as a function of increasing $J_{\pm\pm}$ (color coded) at the corresponding critical field $H_c(J_{\pm\pm})$ in each case. Symbols indicate the gap closing wavevector.}
		\label{fig:SWT_FullH_Condensation}
\end{figure}

The nearest-neighbor dipolar Hamiltonian in Eq.~(\ref{eq:NN_dipolar}) has the above form with $J_{\pm\pm} = -3K_{\mathrm{d,nn}}g_{xx}^2/4 < 0$, and $J_{z\pm}= 0$. The LSWT spectrum for a field $H\| c$ is known analytically \cite{Li2016TLFAM}:
\begin{multline}
	\hbar\omega_{\mathbf{k}} = 2S\sqrt{\mathcal{A}^2 - |\mathcal{B}|^2} , \\
	\mathcal{A} = g_{zz}\mu_{\rm B}\mu_0H/(2S) - 3J_{zz} + \Delta J_{zz} \sum_{i=1}^3 \cos(\mathbf{k}\cdot \mathbf{a}_i), \\
	\mathcal{B} = 2 J_{\pm\pm} [\cos(\mathbf{k}\cdot \mathbf{a}_1) + e^{-i\frac{2\pi}{3}} \cos(\mathbf{k}\cdot \mathbf{a}_2)  + e^{+i\frac{2\pi}{3}} \cos(\mathbf{k}\cdot \mathbf{a}_3)]
\label{eq:TLAFM_Dispersion}
\end{multline}
The dispersion relation $\hbar\omega_{\mathbf{k}}$ is independent of $J_{z\pm}$ as this term does not lead to quadratic terms in the spin wave Hamiltonian, therefore we disregard this term from here onward. Note that the $\mathcal{B}$ term reduces the overall energy of the magnon across the Brillouin zone. For $J_{\pm\pm}=0$ the dispersion surface minimum occurs at the K point. A finite $J_{\pm\pm}$ introduces dispersion modulations near the K point and shifts the minimum away from K, by an amount that depends on the magnitude $|J_{\pm\pm}|$. This fine structure of the dispersion is only apparent very close to the critical field, namely when $|J_{\pm\pm}| \sim g_{zz}\mu_{\rm B}\mu_0 (H - H_{c0})$, with $H_{c0} = 2SJ_{zz}\left(3 + 3\Delta/2\right)/({g_{zz}\mu_{\rm B}\mu_0 )}$ the critical field of the pure XXZ model. This is illustrated in Fig.~\ref{fig:SWT_FullH_Condensation}(a), for $J_{\pm\pm} = 0.08J_{zz}$. Note that at a magnetic field significantly above $H_c$ (top curve) the dispersion fine structure disappears with the minimum restored at the high-symmetry K point. Fig.~\ref{fig:SWT_FullH_Condensation}(b) shows the magnon spectrum calculated from LSWT for different values of $J_{\pm\pm}$. The gap-closure point shifts continuously as $J_{\pm\pm}$ increases. At 15\% of the main exchange, the gap closure point reaches the Brillouin zone mid-edge M point $(1/2,1/2)$. 
 
To obtain a quantitative match with the experimental data, we numerically computed the magnon dispersion relation including the full long-ranged dipolar interactions in the whole triangular layer and adding also a variable nearest-neighbor bond-dependent exchange anisotropy term $J_{\pm\pm}$ (as given in Eq.~(\ref{eq:TLAFM_Full})), adjusted such that the calculated magnon gap closure wavevector $\mathbf{Q}_c$ as the field approaches $H_c$ from above matches the onset wavevector of the magnetic Bragg peaks observed experimentally just below $H_c$, characterized by the incommensuration parameter $\delta_c$ = 0.027 from Fig.~\ref{fig:NBCO_Fan_diffraction}. This gives $J_{\pm\pm} = -1.9\mu$eV, around 1.5\% of the main exchange $J_{zz}$, equivalent to an enhancement of 40\% of the nearest-neighbor part of the dipolar couplings.

\begin{figure}
		\centering
		\includegraphics[scale=1]{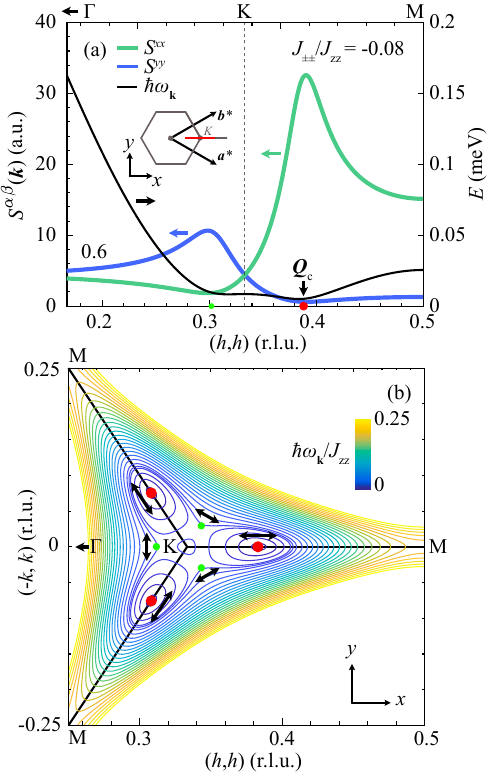}
		\caption{Polarization dependence of the magnon dynamical structure factor just above the polarizing critical field $H_c$. (a) Along the ($h,h$) direction, the magnon dynamical structure factor $S^{\alpha\beta}(\mathbf{k})$ is diagonal in the $xyz$ frame shown in the inset, where the red segment denotes the reciprocal space path in the main panel. As the field approaches $H_c$ from above, $S^{xx}$ (green curve) diverges and $S^{yy}$ (blue) goes to zero at the wavevector $\mathbf{Q}_c$ (vertical black arrow and red dot on the axis) where the magnon dispersion (black solid line, right axis) closes gap at $H_c$. Note how for the wavevector of the shallower local dispersion minimum (green dot on the axis), the roles of $x$ and $y$ axes are interchanged. (b) Contour map of the magnon dispersion relation $\omega_{\mathbf{k}}$ in the vicinity of the K point, at the same field as in (a). Three degenerate global minima are clearly seen (red dots), along each of the three K-M directions. The shallow minima of (a) are indicated by green dots. Black double-headed arrows indicate the polarization of the dominant eigenvalue of the $S^{\alpha\beta}$ matrix as the field approaches $H_c$ from above.}
		\label{fig:SWT_3fold}
\end{figure}	

Next, let us analyze the dynamical spin structure factor just above the critical field. For illustrative purposes, we assume an artificially large bond-dependent anisotropic exchange $J_{\pm\pm} = -0.08J_{zz}$ (with the negative sign to mimic the effect of the nearest-neighbor dipolar couplings, as discussed above) and no dipolar couplings. Bond-dependent terms lead to a non-constant structure factor across the magnon branch. A convenient frame to write the magnon dynamical structure factor $S^{\alpha\beta}(\mathbf{k})$ is the $xyz$ frame shown in the inset of Fig.~\ref{fig:SWT_3fold}(a), where $x$ is along the ($h,h$) direction, $y$ is along ($\bar{k},k$) and $z$ is the out-of-plane direction, which coincides with the applied field direction, note that one of the KM directions is along $x$. Fig.~\ref{fig:SWT_3fold}(a) shows the magnon dynamical structure factor along the ($h,h$) direction. Out of all the components, only $S^{xx}$ and $S^{yy}$ are nonzero. These two components vary strongly with momentum $\mathbf{k}$, becoming equal exactly at the K point, as expected from the three-fold rotational symmetry of the Hamiltonian. However, at the minimum of the dispersion ($\mathbf{Q}_c$, red dot on axis), the contributions of each component are maximally different. $S^{xx}(\mathbf{Q}_c)$ diverges while $S^{yy}(\mathbf{Q}_c)$ goes to zero as $H$ approaches $H_c$ from above, meaning that only the $S_x$ spin components become critical at gap closing. This suggests that a coplanar structure in the $x$-$z$ plane will be stabilized in the phase below gap closing at $H_c$, with spins mostly polarized along $z$ but with a spontaneous incommensurate transverse ordered moment along $x$.  

Note that at the shallower minimum (green circle in Fig.~\ref{fig:SWT_3fold}(a)), the opposite applies - $S^{yy}$ grows as $H\rightarrow H_c^+$, while $S^{xx}$ vanishes.  The overall sign of $J_{\pm\pm}$ does matter in determining which spin component becomes critical. Upon changing the sign of $J_{\pm\pm}$ from $-$ to $+$ the gap closing occurs at the same field and for the same location in reciprocal space, however, the roles of $x$ and $y$ axes swap,  with $S^{yy}$ diverging at gap closure, favoring a coplanar state in the $yz$ plane instead of the $xz$ plane. 

We finally recall the three-fold rotational symmetry expected around the K point. In fact, the magnon branch near the K-point possesses three degenerate minima that appear as $H$ approaches $H_c$ from above, one for each of the equivalent K to M directions. These correspond in our notation to the propagation vectors $\bm{\eta}_1$, $\bm{\eta}_2$, and $\bm{\eta}_3$. Fig.~\ref{fig:SWT_3fold}(b) shows a contour plot of the magnon dispersion relation around the K point for $H\gtrsim H_c$. At each of the minima, the divergence of the appropriate eigenvalue of $S^{\alpha\beta}(\mathbf{k})$ suggests stabilization of a coplanar structure contained in the plane defined by the $z$ axis and the particular $\bm{\eta}_i$ vector (shown by the black double headed arrows in Fig.~\ref{fig:SWT_3fold}(b)). In contrast, the shallower minimum in Fig.~\ref{fig:SWT_3fold}(a) (green dots) corresponds to saddle points of the dispersion in Fig.~\ref{fig:SWT_3fold}(b), where the dominant component of the dynamical structure factor is orthogonal to the corresponding $\bm{\eta}_i$.  

The simultaneous spin-gap closure at several different points raises the question whether the resulting magnetic order is single-$\mathbf{Q}$ or multi-$\mathbf{Q}$. However, our experiments are clear in this regard, and magnetic order is found to be described by multiple magnetic domains, each one single-$\mathbf{Q}$, corresponding to propagation vectors $\bm{\eta}_i^\pm$ with $i=1$, 2, or 3. 

\subsection{The Incommensurate Fan phase}

\begin{figure}
		\centering
		\includegraphics[scale=1]{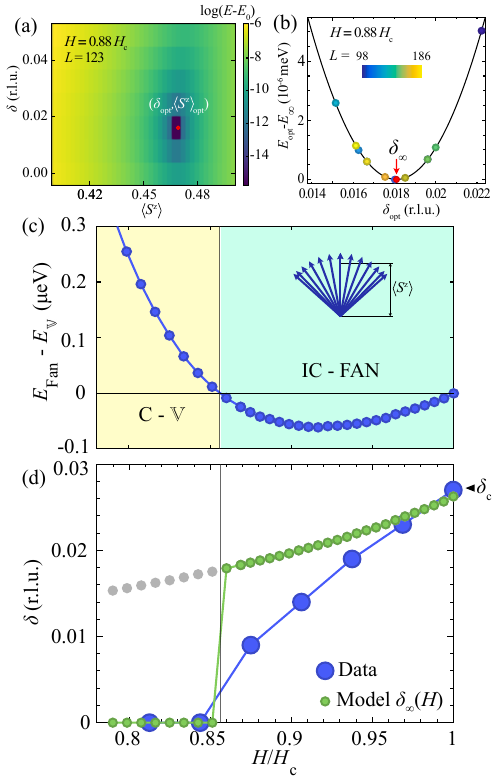}
		\caption{Search of the optimal fan structure at a given field $H$. (a) False colour plot (log scale) of the energy of fan structures constructed as described in the text for a fixed system size $L$, as a function of magnetization $\langle S_z\rangle$ and incommensuration $\delta$, used to find optimal parameters ($\delta_{\rm opt}$, $\left<S_z\right>_{\rm opt}$) (red dot) that give the optimal energy $E_{\rm opt}$. (b) Comparing $E_{\rm opt}$ for different lattice sizes allows for an extrapolation to the thermodynamic limit $L\rightarrow\infty$ assumed to be achieved at the incommensuration $\delta_\infty$ (red vertical arrow) that gives the minimum energy, $E_\infty$. Symbols are color-coded by the $L$ value and the black line is a fit to a quadratic form as described in the text. (c) Comparing the energy of the optimal incommensurate fan structure and the corresponding commensurate $\mathbb{V}$ phase at the same magnetic field, we determine the phase transition $\mathbb{V}\to$ Fan. Note that the Fan is the lower energy structure in the range between $H_{\mathbb{V}-\rm{Fan}}$ and $H_c$. The sketch shows a Fan structure with all spins sharing the same origin. (d) Evolution of the incommensurability of the Fan phase as a function of field (green circles), compared to the experimental values (blue circles) extracted from Fig.~\ref{fig:NBCO_Fan_diffraction}. The model correctly captures the trend of the evolution of $\delta(H)$ and the transition to the phase with commansurate in-plane $\mathbb{V}$ order. Grey circles correspond to optimal fans beyond the limit of phase stability of the fan phase, in the mean-field approach. Magnetic fields for the model are shown normalized to the critical field for the model given in Table~\ref{table:Full_H}, as discussed in the text, while those for data are referenced to the critical field determined from the diffraction data in Fig.~\ref{fig:NBCO_Fan_diffraction}.}
		\label{fig:Fan_model}
\end{figure}	

In the previous sub-section, we argued that magnon gap closure upon approaching the critical field from above is expected to result in the spontaneous onset of a single-$\mathbf{Q}$ incommensurate coplanar magnetic structure below $H_c$. Here we examine in more detail the form of the resulting magnetic structure assuming a classical picture of spins being fixed-length vectors and argue that the pre-saturation magnetic structure is an incommensurate \textit{fan}. In this structure the spin vectors' orientation oscillates sinusoidally in a plane back and forth around the $z$-axis upon moving between lattice sites, with a schematic illustrated in Fig.~\ref{fig:Fan_model}(c) (right inset) with all spins drawn with a common origin. 
Fan structures are commonly found as pre-saturation phases in theoretical studies of Hamiltonians with exchange competition and anisotropy \cite{Fannagamiya1962,Fankitano1964,StarykhTLAFM_Saturation,FanJohnston,FanJohnston2,FanMagnons,IncommJJPrimeBalents}. Such incommensurate spin structures have been observed experimentally in a wide range of systems \cite{HolmiumFan,MnPFan,FanIshikawa1969,FanRbFeMoO4,FanDouble,CsCoBrFan}, for which neutron diffraction has provided key experimental insight in their identification \cite{FanRbFeMoO4,FanMnP_Diff,FanSchobinger1999magnetic,DysprosiumFan,CsCoBrFan, FanRbFeMoO4_diff}.

In our case, we find that the fan is stable for a finite field range from $H_c$ down to $H_{\mathbb{V}-\rm{Fan}}$. Below this, a transition occurs to the commensurate $\mathbb{V}$ phase, which would have been the pre-saturation phase for the pure XXZ model in the absence of bond-dependent anisotropic exchange or dipolar couplings. We also find that in the fan phase the in-plane magnetic propagation vector moves upon reducing field along the Brillouin zone boundary towards the K point, as observed in the experiment. 

Before moving forward, we recall that the experimentally observed magnetic Bragg peaks appear at finite interlayer position $q_z$ indicating a phase offset between the magnetic structure in adjacent layers due to finite inter-layer interactions neglected in our analysis so far. Dipolar interactions of course couple spins in different layers, but we find those alone do not explain the values of $q_z$ seen experimentally, so we consider additional superexchange interactions on various interlayer bonds as illustrated in Fig.~\ref{fig:GroundStates}(b) inset. We defer a detailed discussion of this minimal interlayer exchange model until the following sub-section, and note that in all numerical calculations of ground state energies from this point onwards we use the full Hamiltonian in Eq.~(\ref{eq:Full_H}) with interlayer interactions with parameters in Table~\ref{table:Full_H} and include also the full long-range form of the dipolar couplings for spin pairs both in the same layers and in distinct layers. To simplify the numerical calculations, we choose by construction interlayer exchanges that give  commensurate inter-layer propagation vectors close to the experimental values, so $q_z = 1/3$ in the fan phase, i.e. the magnetic unit cell is commensurate with 3 layers along the vertical direction, to get close to what is seen experimentally in phase IV with $q_z$ incommensurate, but close to 1/3. Upon lowering field we find a transition to a commensurate structure with a $\mathbb{V}$ spin arrangement in each layer with an ABC 3-layer stacking periodicity as illustrated in Fig.~\ref{fig:Structures}(c). Thus, henceforth we identify phase III seen experimentally with this commensurate (C) $\mathbb{V}$ structure, and phase IV with the incommensurate (IC) fan, both with a 3-layer periodicity, and below we consider the C-IC transition between them. 

Unconstrained energy minimization in lattices up to $30\times30\times 3$ sites for fields slightly below $H_c$ converge to coplanar fan phases. However, access to much larger lattices is needed in order to capture the field dependence of the in-plane incommensuration, where energy minimization becomes unreliable. To circumvent this issue, we compare the energy of different fan structures for each magnetic field in order to find the optimal structure. 

We define the fan structure following the convention in \cite{Fankitano1964,FanIshikawa1969,FanZieba2000}, whereby the fixed-length spin at position $\mathbf{R}$ deviates by an angle $\theta_\mathbf{R}$ from the $z$ axis towards the in-plane $x$ axis, given by
\begin{equation}
	\sin{\frac{\theta_\mathbf{R}}{2}} = \frac{\xi}{2}\cos(\mathbf{Q}\cdot\mathbf{R})
	\label{eq:Fan_1}
\end{equation}
where $\xi$ is the amplitude of the fan, which represents the order parameter of this phase, and $\mathbf{Q} = (1/3+\delta,1/3+\delta, 1/3)$. The components of the spin at position $\mathbf{R}$ are then expressed as
\begin{equation}
\begin{aligned}
S_{z,\mathbf{R}}/S &= 1 - \frac{1}{2}\xi^2\cos^2(\mathbf{Q}\cdot\mathbf{R}) \\
S_{x,\mathbf{R}}/S  &= \xi \cos(\mathbf{Q}\cdot\mathbf{R}) - \frac{\xi^3}{8}\cos^3(\mathbf{Q}\cdot\mathbf{R})
\end{aligned}
	\label{eq:Fan_2}
\end{equation}
which shows that the transverse spin component ($S_{x,\mathbf{R}} $) leads to Bragg peaks associated with the incommensurate propagation wavevector $\mathbf{Q}$ (the higher harmonics associated with $3\mathbf{Q}$ are expected to be of order $\xi^4$ weaker). Following this approach, a fan is fully defined by its period set by $\delta$ and the magnetization, i.e. the average $\langle S_z \rangle$ set by the order parameter $\xi$ via Eq.~(\ref{eq:Fan_2}). Note that the longitudinal component ($S_{z,\mathbf{R}} $) has Fourier components only at $2\mathbf{Q}$ and higher even harmonics, and their intensity is expected to be of order $\xi^2$  weaker than that of the main $\mathbf{Q}$ peaks, so unlikely to be observed in experiment.  

With this definition, given a system of size $L\times L$ in plane (see representative in Fig.~\ref{fig:Structures}(d)), periodic boundary conditions fix the accessible values of $\delta$. We use a range of different system sizes to probe a fine grid of reciprocal-space points, with $L$ up to 186 sites. The optimal fan parameters $\delta_{\rm opt}$ and $\left<S_z\right>_{\rm opt}$ at a given field $H$ are obtained as follows:
\begin{enumerate}
\item For a fixed system size set by $L$, we generate hypothetical fan phases for the set of allowed incommensurations $\delta$ and magnetizations $\left<S_z\right>$ using Eqs.~(\ref{eq:Fan_1}) and (\ref{eq:Fan_2}). We compute the mean field energy of each of those structures for the Hamiltonian in Eq.~(\ref{eq:Full_H}) assuming an $L\times L\times 3$ magnetic unit cell, and find the optimal values ($\delta_{\rm opt}$,$\left<S_z\right>_{\rm opt}$)  that give the minimal energy $E_{\rm opt}$ among all generated structures, as shown in Fig.~\ref{fig:Fan_model}(a).

\item This process is repeated for all studied values of $L$.

\item We plot the optimized energies $E_{\rm opt}$ as a function of optimal incommensurations $\delta_{\rm opt}$ and find the incommensuration $\delta_\infty$ that gives the minimum energy $E_\infty$. For a small range of values near $\delta_\infty$ this can be achieved via a fit to a quadratic form $E_{\rm opt}= E_\infty + K (\delta - \delta_\infty)^2$ as illustrated in Fig.~\ref{fig:Fan_model}(b). We identify $\delta_\infty$ with the incommensuration in the thermodynamic limit
\end{enumerate}

We iterate this process at each magnetic field $H$ to obtain the dependence $\delta_\infty(H)$ plotted as green symbols in Fig.~\ref{fig:Fan_model}(d). Having found the optimal fan for each magnetic field we compare its mean-field energy to that of the commensurate $\mathbb{V}$ phase also with 3-layer periodicity at the same field. Fig.~\ref{fig:Fan_model}(c) shows that the fan has the lower energy for an extended pre-saturation field range and below this it gives way to the commensurate $\mathbb{V}$ phase. 

The obtained field-dependence of the incommensuration $\delta_{\infty}(H)$ is plotted in Fig.~\ref{fig:Fan_model}(d) (green filled symbols). Note that, as expected for a continuous fan-to-polarized transition, the limiting value of the incommensuration when the field approaches the critical field $H_c$ from below $\delta_{\infty}(H_c)$ is identical (within numerical accuracy) with the value of the incommensuration $\delta_c$ where the magnon gap closes upon approaching the critical field from above. This provides an important internal consistency cross-check of the numerical method used to determine $\delta_{\infty}(H)$. The calculated incommensuration is predicted to decrease upon reducing field below $H_c$, reproducing qualitatively the trend observed experimentally (blue filled symbols), although experiments observe a much more rapid evolution upon varying field than predicted. Despite this quantitative deviation, the model accurately predicts that there must be a further transition at a lower field to the commensurate $\mathbb{V}$ phase, with the relative transition field $H_{\mathbb{V}-\rm{Fan}}/H_c$ consistent with the experiment.


\subsection{Inter-layer coupling}

\begin{figure}
		\centering
		\includegraphics[scale=1]{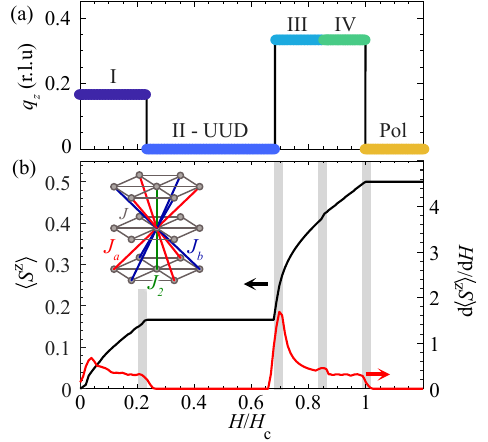}
		\caption{Ground state properties of the full spin Hamiltonian in Eq.~(\ref{eq:Full_H}), as a function of $c$-axis magnetic field (normalized to the critical field, $H_c$, from parameters in Table~\ref{table:Full_H}). (a) Out-of-plane $q_z$ component of the magnetic propagation vector $\mathbf{Q}$. Color coding denotes different magnetic structures (representatives shown in Fig.~\ref{fig:Structures}). Labels I-IV and Pol (polarised) refer to phases observed experimentally. (b) Average spin expectation value along the field direction (black, left axis) and its field derivative (red, right axis). The inset shows the convention for exchange couplings in Eq.~(\ref{eq:Interplane}), with $J_a$ (red bonds) inequivalent to $J_b$ (blue bonds) in the $P\bar{3}$ space group. Vertical gray shaded thick lines show experimental transition fields as determined from location of anomalies in the experimental magnetization curve at 0.08~K, from \cite{sheng2022two} with $\mu_0H_c \sim 1.7$~T.}
		\label{fig:GroundStates}
\end{figure}

\begin{figure}
		\centering
		\includegraphics[scale=1]{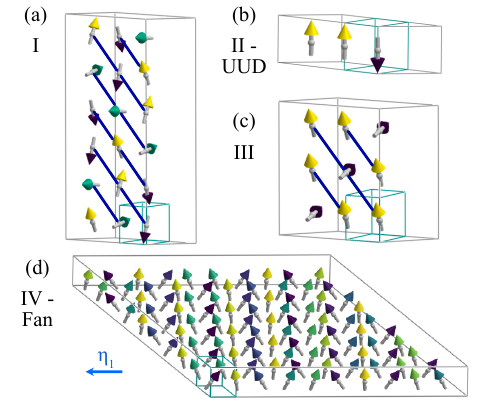}
		\caption{Representative magnetic structures of the full Hamiltonian in Eq.~(\ref{eq:Full_H}) in its four distinct magnetic ordered phases I to IV as a function of increasing magnetic field, as identified in Fig.~\ref{fig:GroundStates}(a). Arrows are spin vecotrs with heads color-coded by the $S_z$ value (distinct color range in each panel). Blue lines show the dominant interlayer exchange $J_b$. Thin gray lines show the outline of the magnetic unit cell, except for the fan structure in (d) where only one layer is shown for clarity, the actual structure has a 3-layer periodicity. A horizontal arrow in (d) shows the in-plane magnetic propagation vector $\bm{\eta}_1$, chosen to generate the representative fan structure.}
		\label{fig:Structures}
\end{figure}

In this sub-section we address the role of inter-layer interactions in determining the complex evolution of the out-of-plane component of the magnetic propagation vector as a function of increasing $c$-axis field illustrated in the diffraction data in Fig.~\ref{fig:Diffraction_YUV}(b,d,f). Given the small scale of these interactions, it is not possible to obtain a set of unique values for interlayer exchanges based on experimental results.  Here we propose a {\em representative} set of inter-layer couplings able to capture the experimental findings while being consistent with the upper bounds placed by INS data \cite{NBCO_Leonie}.

Diffraction data shows that all structures, except for the 1/3$^{\rm rd}$ magnetization plateau phase II, are incommensurate along the inter-layer direction. Optimization of incommensurate structures is a notably complex problem, so to keep the problem tractable, we approximate the out-of-plane component of the propagation vectors in phases I, III and IV to the nearest commensurate values, so we assume $q_z = 1/6$ in phase I and $q_z=1/3$ in phases III and IV.  

We argue below that the following minimal Hamiltonian is able to reproduce the above changes in $q_z$ as a function of field
\begin{equation}
\mathcal{H} = \Sigma_{\rm layers}(\mathcal{H}_{0} + \mathcal{H}_{\rm d} + \mathcal{H}_{\rm bd} + \mathcal{H}_{\rm inter}) 
\label{eq:Full_H}
\end{equation}
with
\begin{equation}
\mathcal{H}_{\rm inter} = \sum_{n = a,b,2} J_n \sum_{\left\langle\langle ij \right\rangle\rangle_n} \Delta_c \left( S_i^xS_j^x + S_i^yS_j^y \right) +  S_i^zS_j^z
\label{eq:Interplane}
\end{equation}
where the interlayer couplings $J_n$ for pairs of spins $\left\langle\langle ij \right\rangle\rangle_n$ in adjacent layers are as defined in Fig.~\ref{fig:GroundStates}(b), all assumed to be of XXZ type with a common exchange anisotropy parameter $\Delta_c$. 

It has been previously proposed that the competition between symmetry-inequivalent $J_a$ and $J_b$ interlayer exchanges \cite{RadaelliRbFeMoO, RadaelliRbFeMoO_2,NBCO_Leonie} favors an incommensurate propagation vector component $q_z$ along the interlayer direction. Assuming $J_a = J_2 = 0$, $J_b>0$ of Heisenberg form ($\Delta_c = 1$) and assuming no dipolar couplings and no bond-dependent exchanges reproduces $q_z = 1/6$ in phase I and $q_z=0$ in phase II, however it predicts $q_z=1/6$ instead of 1/3 in phase III. 

Including dipolar couplings enables more variability in the inter-layer periodicity of the ground state magnetic structure depending on the relative strength of the interlayer exchanges and the dipolar inter-layer couplings. Many sets of parameters were systematically explored for the inter-layer Hamiltonian in Eq.~(\ref{eq:Interplane}) trialing both Heisenberg ($\Delta_c=1$) and XXZ Ising-like forms ($\Delta_c<1$) but only the latter resulted in a consistent solution. Table~\ref{table:Full_H} presents one such set of parameters that can reproduce the $q_z$ values in all distinct phases I to IV in field. This is to be interpreted not as a unique, but as a {\em representative} set of parameters that gives a consistent description. Using this set of parameters we obtain a mean-field critical field $\mu_0H_c^{\rm MF} = 1.86$~T, which we have used to normalize the fields in the theoretical results of Figs.~\ref{fig:Fan_model}, ~\ref{fig:GroundStates}, and \ref{fig:MomentReduction}.


\begin{table}
\caption{Exchange parameters of the full Hamiltonian in Eq.~(\ref{eq:Full_H}), used to compute the results in Figs.~\ref{fig:Fan_model} and \ref{fig:GroundStates}. $^\dagger$As given in Ref.~\cite{NBCO_Leonie}.} 
\begin{tabular}{c c}
\hline \hline 
$J_{zz}^\dagger$ & 0.1225~meV \\ 
$\Delta^\dagger$ & 0.636 \\ 
$J_{\pm\pm}$ & -1.9~$\mu$eV \\ 
$J_{2}$ & 2.8~$\mu$eV \\ 
$J_{a}$ & 0~$\mu$eV \\ 
$J_{b}$ & 7.9~$\mu$eV \\ 
$\Delta_c$ & 0.08 \\ 
\hline 
$g_{xx}^\dagger$, $g_{zz}^\dagger$ & 4.200, 4.716 \\ 
\hline \hline 
\end{tabular} 
\label{table:Full_H}
\end{table}

The obtained magnetic phase diagram as a function of applied field for the above Hamiltonian is illustrated in Fig.~\ref{fig:GroundStates}, which plots in panel (a) the field dependence of the out-of-plane ordering wavevector $q_z$ and in panel (b) the average spin value $\langle S^z \rangle$. At each field the magnetic structure is found numerically by equilibrating the spins assuming a magnetic unit cell of size $3 \times 3\times m$ with $m=6$, 1 and 3 for phases I, II and III, respectively, and using the more elaborate method of finding the optimal incommensurate Fan using $L \times L \times 3$ magnetic unit cells for phase IV, as described in the previous sub-section, with the phase transition fields determined when energies of adjacent phases swap order. Representative structures for each of the four phases I to IV in field are illustrated in Fig.~\ref{fig:Structures}. The transition fields are mostly dictated by the nearest-neighbor coupling constants $J_{zz}$ and $\Delta J_{zz}$. However, as described above the C-IC transition from phase III to IV ($\mathbb{V}$ to Fan) is driven by the strength of the bond-dependent anisotropic exchange $J_{\pm\pm}$ and the dipolar couplings. The calculated magnetization curve in Fig.~\ref{fig:GroundStates}(b) reveals a weak metamagnetic jump at this transition, manifested as a small peak in the first derivative (red curve, right axis).  

As a final check, we verify that the proposed magnitude of the inter-layer couplings is compatible with the upper bounds set by measurements that probe the magnon dispersion in the inter-layer direction in the polarized state. For the considered Hamiltonian parameters, we obtain a calculated magnon dispersion bandwidth along ($00l$) of 4.4~$\mu$eV, well below the resolution limits of all reported INS measurements of the magnon dispersion in the polarized phase.  
			 	
\section{Discussion}
\label{S:discussion}

We note that the transition reported here upon reducing the field from the high-field polarized phase to spontaneous magnetic order with an incommensurate propagation vector moving on the Brillouin zone boundary on the K-M line bears several similarities to the transition observed in zero field upon cooling in the triangular antiferromagnet RbFeCl$_3$ \cite{wada1982incommensurate}. In that case, the spontaneous magnetic order that sets in below the Ne\'{e}l temperature also has an incommensurate in-plane propagation vector located on the Brillouin zone boundary, displaced away from the K point. This propagation vector has been attributed to dipolar couplings \cite{shiba1982incommensurate, suzuki1983microscopic, suzuki1990effects} with a proposed coplanar magnetic structure, analogous to the field-induced fan structure we propose for \NBCO, stabilized by the combination of dipolar couplings and bond-dependent in-plane anisotropic exchange. 

\begin{figure}
		\centering
		\includegraphics[scale=1]{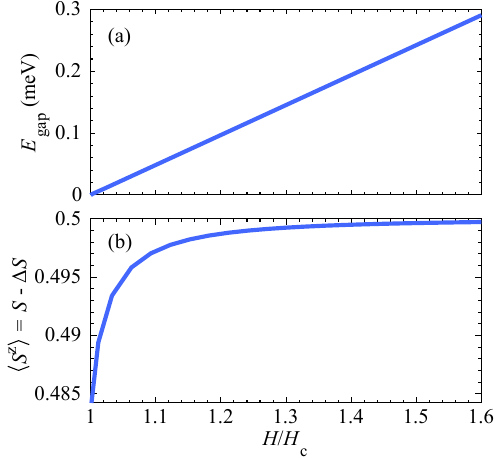}
		\caption{(a) Calculated magnon gap as a function of field in the polarized phase above $H_c$. (b) Magnetization curve for the same field range. $\Delta S$ is the longitudinal moment reduction due to quantum fluctuations.}
\label{fig:MomentReduction}
\end{figure}

For the pure XXZ Hamiltonian in Eq.~(\ref{Eq:IntroH}) full magnetization saturation is achieved exactly at the critical field $H_c$. Quantum fluctuations at higher fields are completely absent because of the continuous rotational symmetry of the Hamiltonian around the field axis, which makes the polarized state an exact eigenstate. However, bond-dependent anisotropic exchange and dipolar couplings break this $U(1)$ symmetry, resulting in enhanced quantum fluctuations at the critical field and the field region above it, with full magnetization saturation expected to be reached only asymptotically upon progressively increasing field above $H_c$. We show in Fig.~\ref{fig:MomentReduction}(b) the numerically obtained magnetization curve for the full Hamiltonian in a LSWT approach using \textsc{sunny}. As expected, the moment reduction due to quantum fluctuations induced by the presence of the small bond-dependent anisotropic exchange and dipolar couplings is very small compared to the full moment value. This implies that quantum fluctuations are also relatively small for fields just below the critical field $H_c$, validating the applicability of the semiclassical, mean-field approach we have used to capture the key properties of the pre-saturation Fan phase in the region immediately just below $H_c$. 

We note however that quantum fluctuations are expected to progressively increase upon further reducing field and at some point the mean-field description may require significant corrections. In particular, our mean-field approach predicts a first-order incommensurate to commensurate transition upon lowering field between the pre-saturation Fan and the lower-field $\mathbb{V}$ phase, see Fig.~\ref{fig:Fan_model}(d) (green symbols). Increasing quantum fluctuations will favour commensurate order, and may contribute to a faster change in $\delta$ and a smooth pinning of the propagation vector to the commensurate K point. This phenomenon has been extensively studied theoretically in the spatially anisotropic Heisenberg triangular antiferromagnet, defined by $J$ and $J^\prime$ couplings on isosceles triangles \cite{StarykhTLAFM_Saturation,starykh2015unusual,IncommJJPrimeBalents}. In this system, any $R = 1-J^\prime/J > 0$ will lead to an incommensurate structure just below $H_c$ (and in particular a planar fan for $R\lesssim0.5$), as a result of magnons condensing at an incommensurate $Q_c$ (the minimum of the dispersion in the polarized state). As the field is reduced, the increasing magnon population leads to enhanced fluctuations that renormalize the magnon-magnon interactions (compared to the bare interaction at $H_c$) favouring commensurate correlations \cite{IncommJJPrimeBalents}. As a result, for small enough $R$, an additional phase transition between the IC fan and a commensurate $\mathbb{V}$ phase is expected at a field value $H_{\mathbb{V}\rightarrow\rm{Fan}}<H_c$ set by the value of $R$. This problem can be cast into a sine-Gordon model, which in 2D predicts a second order C-IC phase transition with a continuous transition of $Q$ from a commensurate position to an incommensurate value, as field increases past $H_{\mathbb{V}\rightarrow\rm{Fan}}$ \cite{ICtoCtransitionBak,pokrovskii1979theory,pokrovskii1980theory, IncommJJPrimeBalents}. The same phenomenology could be likely expected to apply to \NBCO, where exchange anisotropy ($J_{\pm\pm}$ terms) and dipolar coupling play the role of $R$. An account of quantum corrections in the ordered phase is expected to recover a continuous phase transition between the commensurate $\mathbb{V}$ and the Fan as suggested by our experimental data (blue symbols in Fig.~\ref{fig:Fan_model}(d)); however, a formal treatment is beyond the scope of this work.


Finally, a comparison between the presented model and experimental probes is in order. The transitions fields between phase I, II and III are generally consistent with most bulk probe observations, as illustrated by the gray shaded vertical bars in Fig.~\ref{fig:GroundStates}(b), once experimental and theoretical fields are normalised to their respective critical fields, i.e., $\mu_0H_c$ = 1.7~T for experiment and $\mu_0H_c^{\rm MF} = 1.86$~T for calculations.  The difference between the observed $H_c$ and calculated $H_c^{\rm MF}$ of roughly 9\% is not fully understood, but could be due to the simplified nature of the considered Hamiltonian or the fact that we are not optimizing the dominant exchange values from \cite{NBCO_Leonie}. The agreement between experiment and model in the normalised transition fields for transitions I$\to$II, and II$\to$III is not surprising, given that those transition fields are mostly defined by the parameters of the dominant XXZ Hamiltonian, i.e. $J_{zz}$ and $\Delta$, and the additional subleading terms, exchange anisotropy $J_{\pm\pm}$, dipolar and interlayer couplings in Eq.~(\ref{eq:Full_H}) only add small corrections. In contrast, the transition III$\to$IV critically depends on the strength of dipolar couplings and bond-dependent exchange. The fan phase reported here has remained elusive in most experimental studies so far. Previous neutron diffraction studies \cite{sheng2022two, xiang2024giant} reported unusually low critical fields, possibly due to the non-observation of Bragg peaks associated with phase IV. Some reports of magnetization for fields $H\|c$ reveal a peak in the first derivative $dm/dH$ \cite{sheng2022two}, in good agreement with the predicted transition from $\mathbb{V}$ to Fan as shown in Fig.~\ref{fig:GroundStates}(b) (red curve). Finally, a recent NMR study \cite{NMR_NBCO} finds an additional phase between $\mathbb{V}$ and saturation in a field range consistent with our diffraction data.  

Interestingly, the iso-structural material with magnetic Mn$^{2+}$ ions instead of Co$^{2+}$ also displays a pre-saturation phase in-between a $\mathbb{V}$-type phase and field polarized in fields along the easy axis \cite{IOP_NaBaMn,PhysRevB_NaBaMn}. Despite a comprehensive analysis of the magnetically ordered structures in this system \cite{SpinStructure_NaBaMn}, the magnetic structure in this pre-saturation phase has not been addressed in the literature, and it would be interesting to understand whether similar physics to that proposed here could also apply in that case.


	\section{
    Conclusions}
	\label{S:conclusions}
    
In conclusion, we have presented a comprehensive neutron diffraction study of the field-induced magnetic phases in the triangular antiferromagnet \NBCO for fields along the Ising axis. Our data reveal a previously unidentified pre-saturation phase with an incommensurate in-plane magnetic propagation vector. Using a mean-field analysis, we have identified this phase as a planar fan, and highlighted the role of bond-dependent exchange anisotropy and dipolar interactions for its stabilization. We have explained how the fan phase arises naturally in a continuous phase transition via magnon gap closure coming from the high-field polarized phase, and we have reproduced the key experimental finding that the in-plane component of the propagation vector in the fan phase moves along the Brillouin zone boundary upon varying field. Finally, we have proposed a minimal set of interlayer exchange coupling parameters that reproduce the distinct values of the out-of-plane component of the magnetic propagation vector in each of the four magnetically-ordered phases in an applied field. It is expected that this model can be applied to the wider family of glaserite magnets Na$_2$Ba$M$(PO$_4$)$_2$ ($M$ = Co, Ni, Mn), and more generally to other AA-stacked triangular antiferromagnets.

	\section{Acknowledgments}	
D.~F. thanks P.~A. Volkov for helpful comments and R.C. thanks Mike Zhitomirsky for useful discussions and drawing our attention to Ref.~\cite{suzuki1983microscopic}. D.F. and R.C. thank L. Woodland for insightful discussions and related collaboration.  This work was partially supported by the Swiss National Science Foundation, under Grant No. P500PT-222229, and by the European Research Council under the European Union’s Horizon 2020 research and innovation programme Grant Agreement Number 788814 (EQFT). The work at the University of Tennessee was supported by the National Science Foundation through award DMR-2003117. The neutron scattering measurements at the ISIS Facility were supported by a beamtime allocation from the Science and Technology Facilities Council \cite{wish}.

\appendix
\section{Exchange matrix in the absence of mirror planes}	
\label{A}

Here we discuss the general form of symmetry-allowed nearest-neighbor exchange in the case of the actual crystal structure of \NBCO. As discussed in Sec.~\ref{S:theory}A, for an idealized crystal structure with vertical mirror planes bisecting the Co-Co bonds the general form of the nearest-neighbor exchange is listed in Eqs.~(\ref{Eq:IntroH}) and (\ref{eq:TLAFM_Full}) with four independent parameters $(J_{zz},\Delta,J_{\pm\pm},J_{z\pm})$  all real. This applies in the case when each bond has a $2/m$ point group, with all bonds being symmetry related by combinations of $3$-fold rotations, site inversion and lattice translations. In the actual crystal structure of \NBCO (space group $P\bar{3}$) the vertical mirror planes are broken by the rotation of the CoO$_6$ octahedra around the $c$-axis, see top-right corner of Fig.~\ref{fig:Introduction}(b). The point group for each bond is reduced to $\bar{1}$ and this lowering of symmetry allows additional exchange terms. Since this case has not been discussed in the literature to the best of our knowledge, we provide below the corresponding generalization of the model in Eq.~(\ref{eq:TLAFM_Full}).


Since inversion symmetry at the center of the bond is still present, any antisymmetric (Dzyaloshinskii-Moriya) exchange is still forbidden, but the most general coupling is a symmetric matrix with six independent parameters. The coupling for the $\bm{a}_1$ bond expressed in the $xyz$ frame has the form
\begin{equation}
\mathcal{H}^{P\bar{3}}_{\mathbf{a}_1} =
\begin{pmatrix}
S_i^x \,\, S_i^y \,\, S_i^z\\
\end{pmatrix}
\begin{pmatrix}
J_{xx} & J_{xy} & J_{xz}\\
J_{xy} & J_{yy} & J_{yx} \\
J_{xz} & J_{yz} & J_{zz} \\
\end{pmatrix}
\begin{pmatrix}
S_j^x\\
S_j^y\\
S_j^z\\
\end{pmatrix}
\label{eq:NN_dipolar_147_mat}
\end{equation}
Interestingly, using the common crystallographic XXZ-like notation we may bring the general coupling into a very similar form to Eq.~(\ref{eq:TLAFM_Full}) by allowing the pseudodipolar parameters $J_{\pm\pm}$ and $J_{z\pm}$ to be \textit{complex}. This clearly extends the number of independent parameters from four to six and the nearest neighbor Hamiltonian from Eq.~(\ref{eq:TLAFM_Full}) generalizes to 
\begin{multline}
\mathcal H_{1}^{P\bar{3}}=\sum_{\langle ij\rangle}J_{zz}\big[S_i^zS_j^z+\Delta\big(S_i^xS_j^x+S_i^yS_j^y\big)\big]+ \\
\big(J_{\pm\pm}\gamma_{ij}S_i^+S_j^+ + J_{\pm\pm}^*\gamma_{ij}^*S_i^-S_j^-\big) \\
-\frac{i}{2}\Big[\big(J_{z\pm}\gamma_{ij}^*S_i^+-J_{z\pm}^*\gamma_{ij}S_i^-\big)S_j^z+S_i^z\big(J_{z\pm}\gamma_{ij}^*S_j^+-J_{z\pm}^*\gamma_{ij}S_j^-\big)\Big]
\label{eq:NN_dipolar_147}
\end{multline}
with $\gamma_{ij} = 1,  e^{i \frac{2\pi}{3}}$ and $ e^{-i \frac{2\pi}{3}}$ for bonds along $\mathbf{a}_1$, $\mathbf{a}_2$ and $\mathbf{a}_3$, respectively, as defined in Fig.~\ref{fig:SWT_NN_Dip}(a). Taking $J_{\pm\pm}$ and $J_{z\pm}$ as complex allows us to express the parameters in Eq.~(\ref{eq:NN_dipolar_147_mat}) in a simple form, as
\begin{multline}
 J_{xx}=\Delta J_{zz}+2\mathrm{Re}(J_{\pm\pm}), \quad\quad J_{yy}=\Delta J_{zz}-2\mathrm{Re}(J_{\pm\pm}),\\
 J_{yz}=\mathrm{Re}(J_{z\pm}), \\
 J_{xz}=\mathrm{Im}(J_{z\pm}),\quad\quad J_{xy}=-2\mathrm{Im}(J_{\pm\pm})
\label{eq:NN_dipolar_147_terms}
\end{multline}

While the terms associated with $J_{z\pm}$ (even in its complex form) do not enter the LSWT dispersion for spins polarized along the crystallographic $c$ axis, both the real and imaginary components of $J_{\pm\pm}$ do contribute. However, we stress that the nearest-neighbor part of the dipolar interaction considered in the main text can be written in the form of Eq.~(\ref{eq:TLAFM_Full}), and so the imaginary part of $J_{\pm\pm}$ is disregarded in the main text.

We finally we note that imposing 2-fold symmetry of rotation around the axis of the bond (or mirror plane bisecting the bond) forces both $J_{\pm\pm}$ and $J_{z\pm}$ to become real, by mapping $S^+$ to $S^-$ and $S^z$ to $-S^z$. Applying these transformations to Eq.~(\ref{eq:NN_dipolar_147_mat}) and imposing symmetry results in $J_{\pm\pm} = J_{\pm\pm}^*$ and $J_{z\pm}=J_{z\pm}^*$, thus recovering the exchange model in Eq.~(\ref{eq:TLAFM_Full}). 

\bibliography{PRB23_v1.bib}
\end{document}